\documentclass[12pt]{article}
\usepackage{amsmath}
\usepackage[english]{babel}
\usepackage[applemac]{inputenc} 
\usepackage[T1]{fontenc}       
\usepackage{graphicx}
\usepackage{cite}
\usepackage[left=2cm,right=2cm]{geometry}
\begin{document}

\title{Will Trump win again in the 2020 election? \\
An answer from a sociophysics model}

\author{Serge Galam\thanks{serge.galam@sciencespo.fr} \\ CEVIPOF - Centre for Political Science, \\ Sciences Po and CNRS,\\ 98 rue de l'Universit\'e Paris, 75007, France}
 
\date{}

\maketitle

\abstract{ This paper predicting Trump victory has been submitted before the election and revised after, allowing to add a Foreword and Note Added in Revision to discuss in details the causes of the failure of the prediction.\\

In 2016, Trump was unanimously seen as the loser in the November 8 election. In contrast, using a model of opinion dynamics I have been developing for a few decades within the framework of sociophysics, I predicted his victory against all odds. According to the model, the winning paradoxical martingale of 2016, has been Trump capability to activate frozen prejudices in many voters by provoking their real indignation. However, four year later, Trump “shocking” outings do not shock anymore, they became devitalized, losing their ability to generate major emotional reactions. Does this mean that this time around he will lose the 2020 election against Biden, as nearly all analysts, pundits and commentators still predict? No, because with frozen prejudices remaining frozen, the spontaneous prejudices will be activated but this time they will benefit to both Biden and Trump. The main ones are the fear of the other candidate policy and the personal stand facing a danger. In addition, Trump presidency having polarized a large part of American voters into narrow-minded anti-Trump and narrow-minded pro-Trump, those I denote in my model as inflexibles, will be driving the dynamics of choices. Both effects, prejudices and inflexibles can either compete or cooperate making their local combination within each state, decisive to determine the faith of the state election. As a result, tiny differences can make the outcome. Based on my rough estimates of  associated proportions of inflexibles and prejudices, the model predicts Trump victory in the 2020 November election.

}

\section*{Foreword}

Applying the Galam model of opinion dynamics to predict the outcome of the November 3, 2020 US presidential election, I concluded on Trump victory. While the paper was logically submitted prior to the election, it happened that the report by the referees came once the outcome of the election has been known: my prediction had failed. On this basis, I could have been tempted to withdraw the paper, no researcher being eager to have in print a wrong prediction. 

Nevertheless, as clearly stated in the conclusion of the first version of the submitted paper, making predictions is what matters to build a hard science approach to model political events. The process is not a one time shot but relies on right and wrong predictions, both being meaningful and instrumental to allow
a series of  back and forth between the model and reality. This is why,  I kept on the submission with a revised version being aware I was making a wrong prediction.

On this basis, to preserve the hard science approach underlying above stand, I have chosen  to keep the paper original version with its wrong prediction having at its end a Note added in Revision where I discuss in details what went wrong and what has been robust in the making of the prediction.

\section{Introduction}

Dealing with the 2020 American presidential election with the incumbent republican candidate Donald Trump being challenged by the democrat Joe Biden, it is of a central importance to go back four years ago when Trump won the 2016 presidential election defeating Hillary Clinton against all odds. 

Then, till 2016, November 8, Trump was predicted to lose the election by almost every analyst, scholar, pundit and polls. Even beforehand, most of them were labelling Trump as a political bubble, which was set to collapse during the Republican nomination campaign, anticipating he would even not last till the end of the process of Republican nomination.

In contrast,  using a model of opinion dynamics \cite{bra} I have been developing for a few decades within sociophysics \cite{bra,noor,fran,sprin,cast}, I did predict Trump victory  \cite{tru}. It is worth to notice that I myself did not believe my prediction could be right but the model was yielding Trump victory. More precisely, I did not predict his victory, I showed why keeping on infuriating millions of American was a winning strategy for him. And indeed he kept on with his repeated shocking statements and he eventually won.

The prediction was based on both the effect of prejudices in the dynamics of opinion and Trump's capacity to modify the hierarchy of hidden prejudices to activate the one favoring him, which otherwise were frozen among most voters. In particular, it has been his repeated shocking statements, which have put ahead frozen prejudices many voters have.

I should  mention that Allan Lichtman, an historian from American University in Washington, did also predicted 2016 Trump victory using a binary scheme of 13 keys  \cite{lich}. 

For the 2020 Presidential campaign, again, most media, analysts, scholars, pundits and polls have kept assessing Trump would lose the election \cite{pol}. Are they right this time?

Indeed, during the last four years they have been anticipating Trump will not complete his mandate. They also started prophesying an impeachment, convinced it will lead to Trump eviction. The impeachment was eventually implemented  by the Democrats at the Congress but without the expected destitution of Trump with the Republicans clearing him at the Senate. 

Democrats and anti-Trump  analysts have been blind to the reality entrenched within wishful thinking with the underlying conviction that Trump “stole” the 2016 victory. Accordingly, not much effort has been done to understand how and why he did win the election. And here we are with Trump running for a second mandate.

Emblematic of this celebrating an anticipated Trump defeat has been the coverage on August 5, 2020, by the New York Time of a video by Lichtman predicting this time, Trump being defeated \cite{lich}.  
As to exorcise last election “curse”, the journalists presenting the video, wrote: “Right now, polls say Joe Biden has a healthy lead over President Trump. But we’ve been here before (cue 2016), and the polls were, frankly, wrong. One man, however, was not. The historian Allan Lichtman was the lonely forecaster who predicted Mr. Trump’s victory in 2016” \cite{lich}. The unwritten conclusion being this time Game is over, Trump will lose the election. Yet, the whole media coverage has a taste of “déjà vu” from the 2016 campaign, which ended with Trump victory. 

What about the model I used to make the 2016 prediction?  After four year Trump “shocking” outings do not shock anymore, they lost their ability to generate emotional reactions turning devitalized. Does this mean that this time around he will lose the 2020 election against Biden? 

No, because with frozen prejudices remaining frozen, the spontaneously  prejudices will be activated but this time they will benefit to both Biden and Trump. The main ones are the fear of the other candidate policy and the personal stand facing a danger. 

In addition, during his mandate Trump has  polarized a large part of American voters into narrow-minded anti-Trump and narrow-minded pro-Trump, those I designate in my model as inflexibles \cite{frans}. And the model shows that inflexibles have a drastic effect on the driving of opinion dynamics. Numerous works have investigated the role of inflexibles \cite{mas,JP18,PSL12,BRG16,CO15,Mo15,MG13,pub}.

Both effects, prejudices and inflexibles can either compete or cooperate making their  local combination within each state, decisive to determine the faith of the state election. Tiny differences can make the outcome \cite{pair}. Based on my rough estimates of associated proportions of inflexibles and distribution of active prejudices, the model predicts Trump victory in the 2020 November election.

The rest of the paper is organized as follows. Section 2 contains a few words of caution about the positioning of the work. The Galam model of opinion dynamics is presented in Section 3 with a review of the local update among a group of discussing agents, the prejudice driven tie breaking effect, the inflexible effect, and the update equation. Section 4 is about the 2016 prediction. The winning strategies for the 2020 election are elaborated in Section 6. The prediction for the 2020 election winner is given in Section 7. The possible role of hidden voting and hidden abstention is briefly mentioned. Concluding statements are made in Section 8.

\section{Words of caution}

It is of importance to emphasize that I am not dealing with a choice being wrong or right. I am not advocating for one candidate or the other. Within the field of sociophysics, I am focusing on identifying the hidden mechanisms, which drive the dynamics of opinion between two competing choices, in particular to anticipate the one, which will eventually ends up above 50\% in case of a vote.

It is worth to remind that sociophysics is the use of concepts and techniques from Statistical Physics to describe some social and political behaviors. It aims neither at an exact description of the reality nor to substitute to social sciences but to provide an additional different and rather counter intuitive vision of the social reality \cite{sprin}. One main topic of sociophysics is opinion dynamics.

\section{The Galam model of opinion dynamics}

Two choices A and B are competing among agents like for a Presidential race (Clinton and Trump in 2016,  Trump and Biden in 2020). I consider heterogeneous agents with three psychological traits, floaters, inflexibles and contrarians.
\begin{itemize}
\item 
Floaters are agents having an opinion and advocating for it but they are susceptible to shift opinion if given convincing arguments
\item
Inflexibles are agents (stubborn, committed) who never shift its opinion.
\item
Contrarian are agents taking a contrary choice to the (local or global) majority. They are not included in the present work.
\end{itemize}

The dynamics of opinion is initiated at time $t_0$ with each agent having made a choice, either A or B. I do not investigate what mechanisms lead to these respective individual choices. I only consider the initial proportions $p_0$ and $(1-p_0)$ of support for respectively A and B at time $t_0$. External events, which can act directly on single individuals to influence at diverse degrees their current choice are included in the making of $p_0$. However, at time $t_0$ all external influences are cut off. The value of $p_0$ can be obtained from polls.

I then model the dynamics of individual shifts of opinions driven by informal discussions among small groups of people arguing about choices A and B during the on going campaign till the voting day. In case a major external event does occur, it will impact directly individual choices with a rescaling of the respective supports thus creating a new initial state. When that happens, a new measure of a new initial value $p_0$ is performed and the dynamics is reactivated. 

\subsection{The local update among a group of discussing agents}

The cognitive and psycho-sociological processes leading one person to shift opinion while discussing an issue informally in a small group of agents, are complicated and unknown. I make it simple, the ``physicist way" with one person one vote plus a local majority rule in the discussing group.

Applying a local majority rule creates a local group polarization. It is worth to note that while stubborn agents do vote as floaters do, contrary to floaters, in case they are minority, they do not follow the majority keeping on their initial individual choices.

The dynamics is implemented by first randomly distribute all agents in a series of small groups of various sizes ranging from 1 to L, where L is rarely larger than 6. Second, all groups are updated according to majority rules. A new value of A support $p_1$ is obtained. Third, groups are dispersed and all agents are reshuffled. The three precedent steps are then iterated yielding,
\begin{equation}
p_0 \rightarrow p_1 \rightarrow p_2  \rightarrow p_3 \rightarrow \dots \rightarrow p_n  ,
\label{dyn}
\end{equation}
where $n$ corresponds to voting time $t_n$. Given $p_0$, I determine if $p_n > \frac{1}{2}$ (A winning the election) or  $p_n < \frac{1}{2}$ (A losing the election). In the last weeks before the vote, the campaign intensifies with people discussing more and more often, which means more updates for a given time duration.

\subsection{The prejudice driven tie breaking effect}

While majority rule always applies for odd size groups, it cannot operate for an even size group at a tie. In such a case, a physicist would likely decide to keep the tie and reshuffle the agents to preserve the symmetry between A and B. 

However, here we are dealing with humans and at a tie, the collective confrontation of individual opposite views gets trapped in a balanced state with both choices getting equally choosable.  A tie creates a collective doubt where rationality cannot operate to make a choice between A and B. This assertion sounds counterintuitive since a priori, aggregating more arguments is expected to provide a more rationalized choice. However, at a tie, aggregation of information neutralizes the information content. Accordingly, the choice is made “randomly” as with flipping a coin with no reason given for the choice made.

But at this moment, I make the hypothesis that indeed the ``flipping coin" is biased along the prejudices activated spontaneously and unconsciously by the issue at stake. The tie is resolved at the benefit of the choice which is naturally in tune with the prejudiced activated by the issue at stake but stays perceived as chosen by chance.

To implement above tie breaking requires identifying the prejudices actually activated to rise the local doubt since prejudices of a given social group are numerous and diverse. Moreover, people are unaware of most of them them. Examples of prejudices are sexism, homophobia, racism, religious beliefs, precaution principe, societal vision. 

In addition, different issues rise different prejudices among the same social people. For instance, a group at a tie dealing with a reform proposal, does chose to reject the proposal driven by a natural ``Tip to the Status Quo". In contrast, the very same group at a tie about choosing a new high tech product, naturally chooses the new product guided by a natural ``Tip to the Novelty". 

\begin{figure}
\begin{center}
\includegraphics[width=8cm,height=6cm]{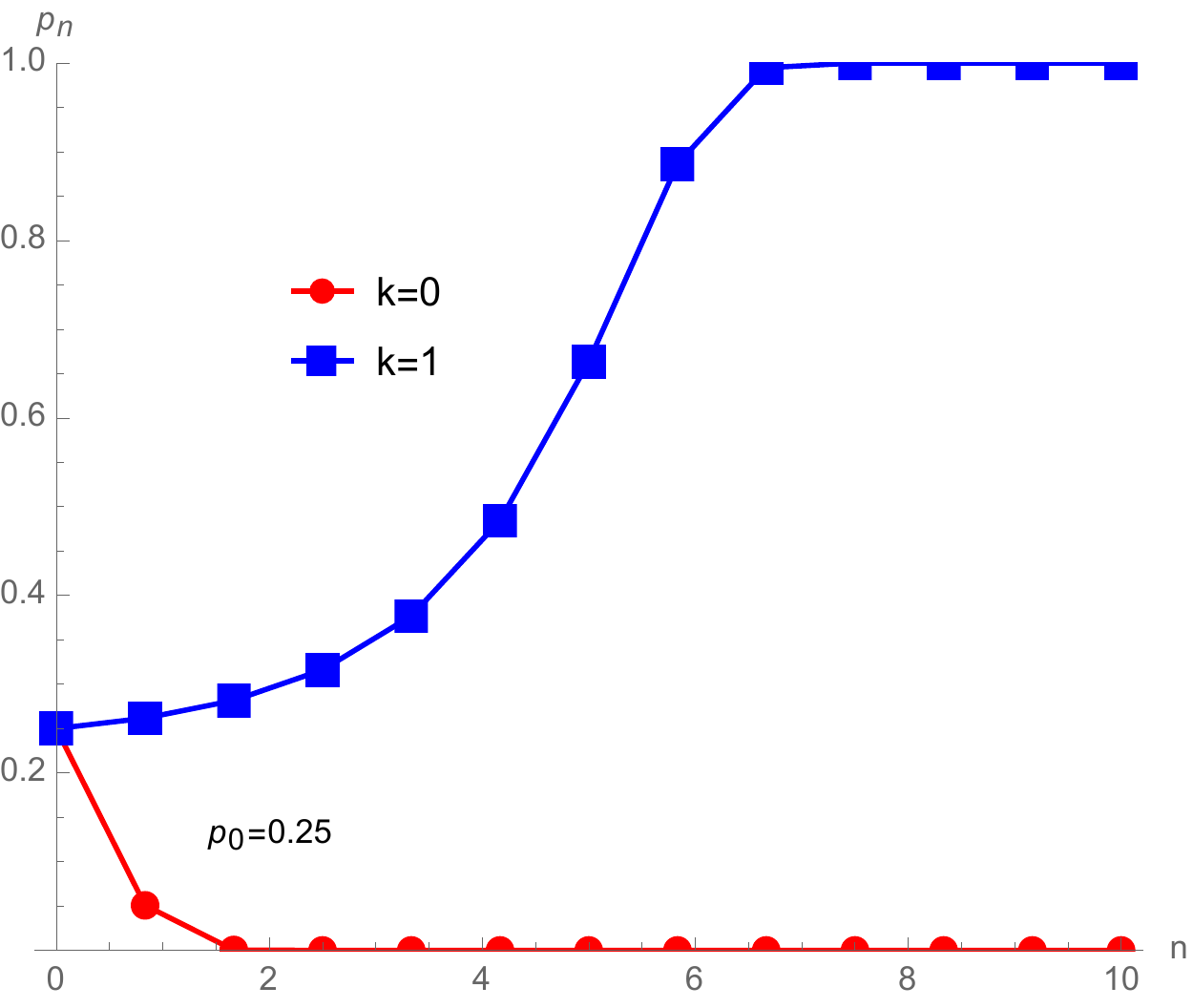}
\includegraphics[width=8cm,height=6cm]{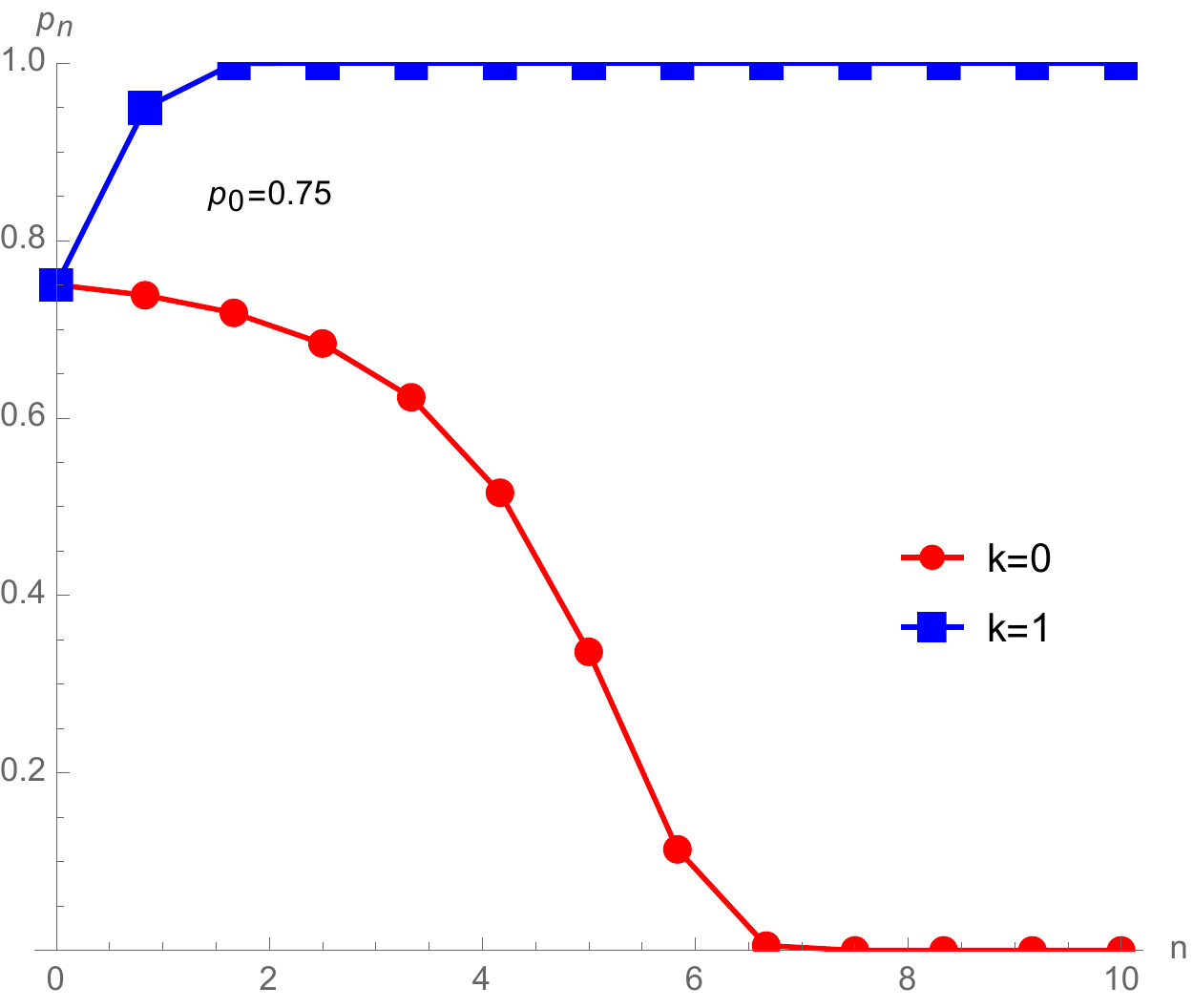}
\caption{Iterations of Eq. (\ref{r4}) for $p_0=0.25$ (left) and $p_0=0.75$ (right) with $k=0$ and $k=1$.
}
\label{r4k}
\end{center}
\end{figure}

For an heterogeneous population, to account for a distribution of different prejudices, the tie breaking goes with a probability $k$ for choice A and $(1-k)$ for choice with $0\leq k \leq 1$ \cite{hete}. For instance, groups of size 4 yield the update equation,
\begin{equation}
p_{n+1} = p_n^4 + 4 p_n^3 (1 - p_n)+ 6 k p_n^2 (1 - p_n)^2 ,
\label{r4}
\end{equation}
whose iterations are shown in Figure (\ref{r4k}) for $p_0=0.25$ and $p_0=0.75$ with $k=0$ and $k=1$.

\subsection{The inflexible effect}

An inflexible does take part in the local making of the majority but does not shift in case its choice is minority in the group. It is worth to stress that an inflexible does not contribute more than a floater to the local majority, every agent having a single vote \cite{frans}. 

Noting $a$ and $b$ the respective proportions of inflexibles holding opinion A and B, the proportions of floaters become $(p_0-a)$ for A and $(1-p_0-b)$ for B. While the dynamics modifies $p_0$, $a$ and $b$ are fixed and do not change during the campaign.The values $a$ and $b$ satisfy the constraints $0 \leq a \leq 1$, $0 \leq b \leq  1$, $0 \leq  a + b \leq  1$. 

\begin{figure}
\begin{center}
\includegraphics[width=8cm,height=6cm]{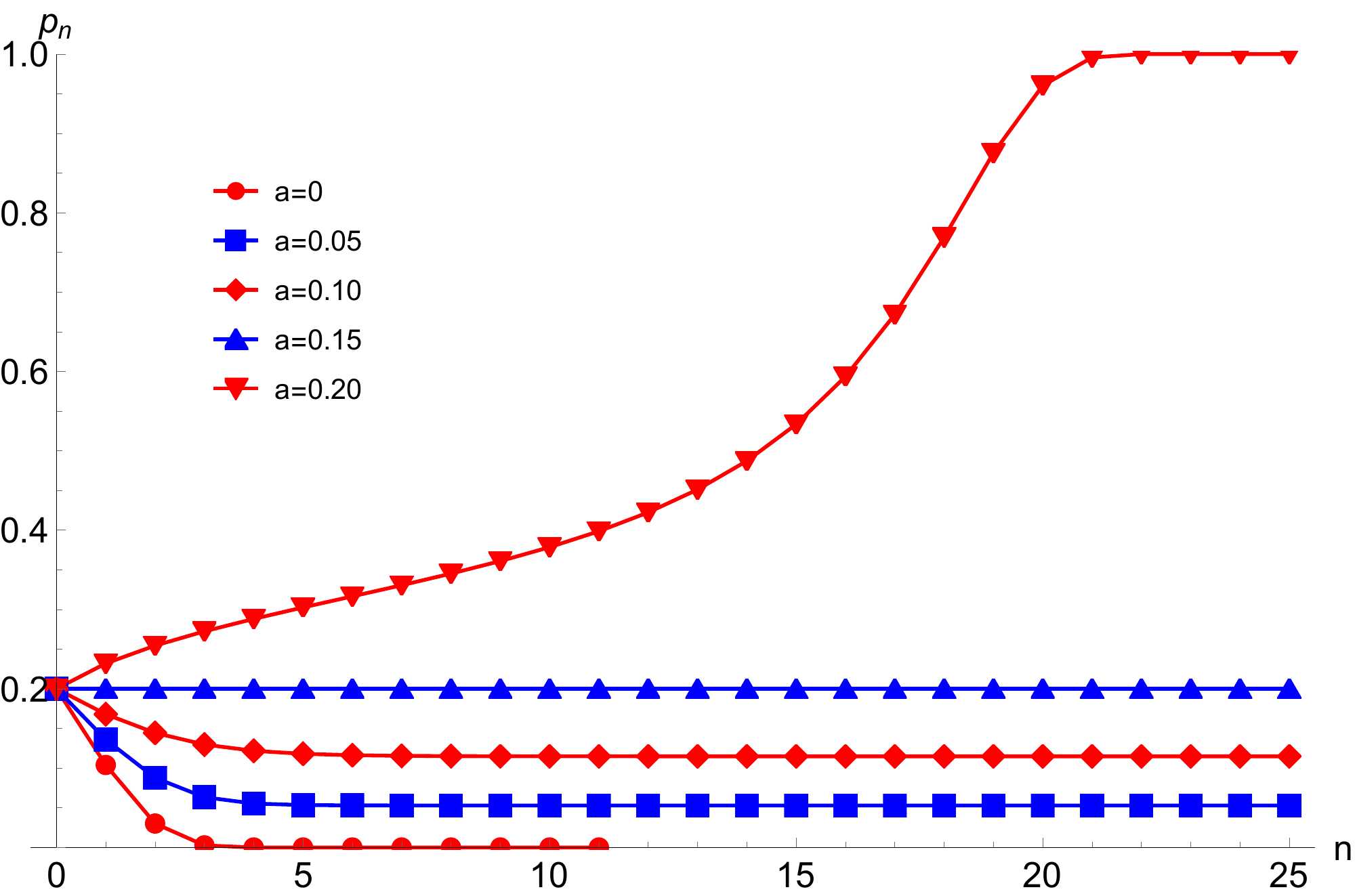}
\caption{Iterations of Eq. (\ref{r3}) for $p_0=0.20$  with $a=0, 0.05, 0.15, 0.15, 0.20$.
}
\label{rr3}
\end{center}
\end{figure}

Figure (\ref{rr3}) illustrates the inflexible effect for the case of one sided inflexibles ($b=0$) with groups of size 3. Starting from $p_0=0.20$ with respectively $a=0, 0.05, 0.15, 0.15, 0.20$,  the associated update equation,
\begin{equation}
p_{n+1} = p_n^3+3p_n^2(1-p_n)+a(1-p_n)^2 ,
\label{r3}
\end{equation}
is iterated 25 times in a row.

\subsection{The update equation}

Previously, I could study the combination of tie breaking and inflexible effects only for group size 2 \cite{pair}. For size 3 it was only inflexibles (Eq. (\ref{r3})) and  tie breaking only for size 4 (Eq. (\ref{r4})). 

However, recently with my colleague Cheon we have been able to derive analytically a universal update equation in a five dimensional parameter space  $(k, a, b, c, r)$ where $c$ is the proportion of contrarians (not included in this study: $c=0$) and $r$ is the group size \cite{uni}. On this basis, it is possible to have the analytical expression of the update equation with any combination and distribution of sizes. However, solving the associated fixed-point equation will very difficult being deployed a high dimensional phase space. To make the analysis simpler and yet incorporate local majority, tie breaking prejudice and inflexible effects, I choose here to have all discussing groups with a size 4. The associated update equation writes, 

\begin{equation}
p_{n+1} = p_n^4 + 4 p_n^3 (1 - p_n)+ 3 a (1 - k) p_n (1 - p_n)^2  - 3 b k p_n^2 (1 - p_n)  + 6 k p_n^2 (1 - p_n)^2  + a (1 - p_n)^3 - b p_n^3   ,
\label{p1}
\end{equation}
which yields the series in Eq. (\ref{dyn}) by repeated iterations.

Solving the fixed point equation $p_{n+1}=p_n$ reveals the dynamics. Two scenarios are obtained, either one tipping point $p_{t}$ located between two attractors $p_A>\frac{1}{2}$ and $p_B<\frac{1}{2}$ or one single attractor $p_s$. In the first case, $p_0 >p_t$ leads to reach $p_A$, i.e., A wins the election while  $p_0 <p_t$ lead to reach $p_B$, i.e., A loses the election. Note that $p_{t}$ can be located above or below $\frac{1}{2}$ depending on $k, a, b$. The second case leads to $p_s$ whatever is the initial value $p_0$ with A winning for $p_s>\frac{1}{2}$ and A losing when $p_s<\frac{1}{2}$.

The tie breaking prejudice effect produces only first scenario with $p_A=1$, $p_B=0$ and $0.23\leq p_t\leq 0.77$ as a function of $k$. In contrast, inflexibles produce the two scenarios depending on two critical values $x_{1}>0$ and $x_{2}<0$ with $x\equiv a-b$. In the range $x_{2} \leq x \leq x_{1}$ the tipping point dynamics prevails while for $x < x_{2}$ or $x >x_{1}$ the dynamics is monitored by one single attractor. First case has $p_A\neq1$ and $p_B\neq 0$ and second one has $a < p_s < 1-b$. 

In the present study, I incorporate both  tie breaking and inflexible effects which can combine to favor the same choice or compete each one favoring a different choice. The tie breaking effect is smooth as a function of $k$ but the inflexible effect is much more drastic and non linear as a function of $x$.

\section{The 2016 prediction}

With respect to 2016 my claim is that Trump victory was neither an accident nor the result of some manipulations. It was the outcome of a non-linear dynamics, which obeys quantitative laws. And indeed, I could predict Trump victory using  the Galam model of opinion dynamics \cite{tru}, which could answer the following questions:
\begin{description}
\item[(i)] 
How comes Trump won while making repeated shocking statements, which infuriated millions of people?
\item[(ii)] 
How comes Trump campaign, which went against all making sense principles has turned successful?
\end{description}

During the 2016 earlier campaign, the  prejudices which were naturally activated in case of a tie, included among others: Trump is not shaped to be President, Trump has not political knowledge for foreign affairs, Trump does not fit to the job, Trump has no political experience. All implied $k = 0$ for Trump, which locates his $p_t$ very high. Accordingly, even with an initial high $p_0$, his support was doomed to shrink. On this basis, applying the model to Trump led me join the overwhelming shared conclusion that he will be defeated during the primary campaign.

But at the beginning of March 2016 I had the chance to be in the US for a conference and I realized that Trump was ``playing" with my model along a novel path I never envision before. While I first thought that prejudices are given and cannot be modified on a short time scale, Trump was innovating with an original scheme to modify the activated prejudices.

Indeed,Tump shocking statements were infuriating many people pushing them to initiate more debates to condemn his statements. But at the same time, this emotional reaction was unfreezing deep locked prejudices, which were present in many of those infuriated voters.
The emotion to condemn the statements brought in front line frozen prejudices, which become the ones  activated at a tie.

\begin{figure}
\begin{center}
\includegraphics[width=8cm,height=6cm]{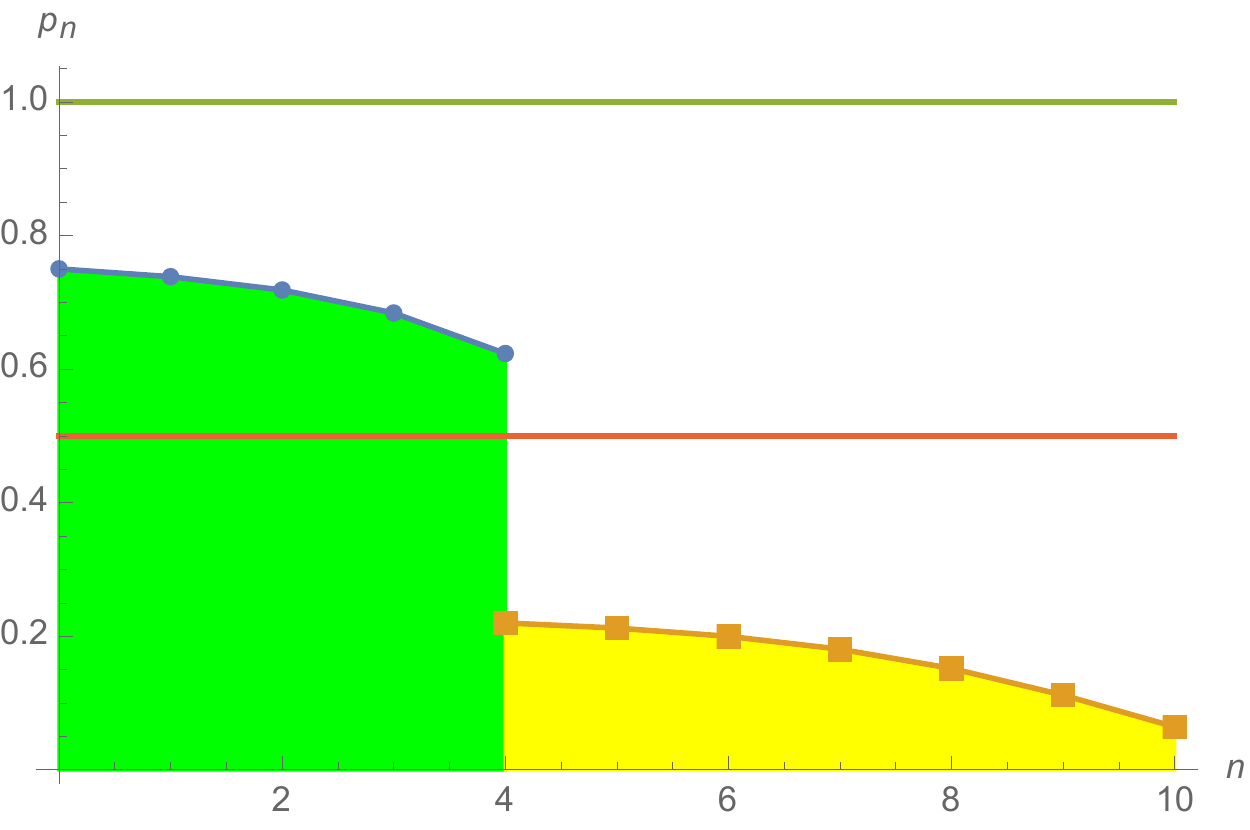}
\includegraphics[width=8cm,height=6cm]{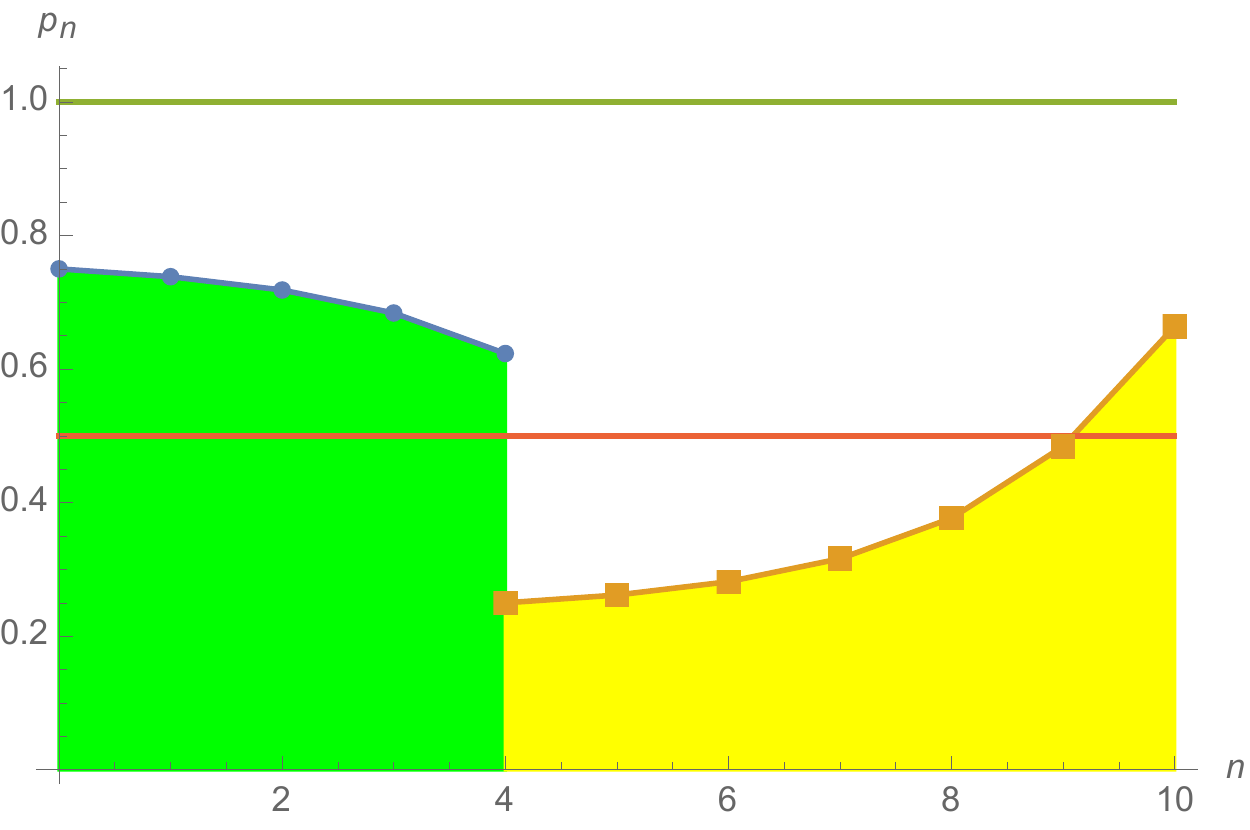}
\caption{Iterations of Eq. (\ref{r4}) with first $p_{0,1}=0.75<p_t \approx 0.77 ( k=0)$ followed respectively by (left) $p_{0,2}=0.22<p_t\approx 0.23 (k=1)$ with Trump losing and (right) $p_{0,2}=0.24 > p_t\approx 0.23 (k=1)$ with Trump winning.
}
\label{u}
\end{center}
\end{figure}

However, the process is twofold. First Trump was losing votes, and second he was turning the tie breaking at his  the benefit.  Therefore, to have the follow up debate increases his support, the starting new support had to be located above the new tipping point created by the change of the activated prejudice. As a result, to win in a given state required both the existence of a minority of openly prejudiced people to ensure the new $p_0$ to be above the new $p_t$ and a substantial proportion of agents sharing the frozen prejudice. Figure (\ref{u}) exhibits the two cases with first $p_{0,1}=0.75<p_t \approx 0.77 (k=0)$ followed respectively by $p_{0,2}=0.22<p_t\approx 0.23 (k=1)$ with Trump losing and $p_{0,2}=0.24 > p_t\approx 0.23 (k=1)$ with Trump winning.

\section{The 2020 prediction: setting}

After four years with Trump president, people got used to his repeated shocking statements, which stopped generating indignation, turning the frozen prejudice effect obsolete for the 2020 campaign. Does that means Trump will be losing the election? The answer is no. 

With no unfreezing mechanism, the naturally activated prejudices  will determine the tie breaking. However, this time, the activated prejudices are activated at the benefit of both candidates. Main ones are fear of the other candidate with either fear for a second Trump term or fear for socialism and chaos. Depending on the individuals, one fear will be more present than the other.

In addition, Trump presidency has created a very high level of polarization among American voters with millions of both stubborn anti-Trump and stubborn pro-Trump voters. Alike the natural prejudice effect, the inflexible effect is available on both sides.

Therefore, what matters is the differences in both the distance from $\frac{1}{2}$ for $k$ and the respective proportions of inflexibles. To have a more immediate reading of the various cases, in addition to  $x\equiv a-b$, I introduce $dk\equiv k-\frac{1}{2}$. Then, $dk>0$ means a prejudice advantage to A and $dk<0$ a prejudice advantage to B with $-\frac{1}{2}\leq dk \leq \frac{1}{2}$. For inflexibles,  $x>0$ means more inflexibles for A than for B and $x<0$ less inflexibles for A than for B.  The constraints on $a$ and $b$ yield $2a-1\leq x \leq 2a$.

\begin{figure}
\begin{center}
\includegraphics[width=8cm,height=8cm]{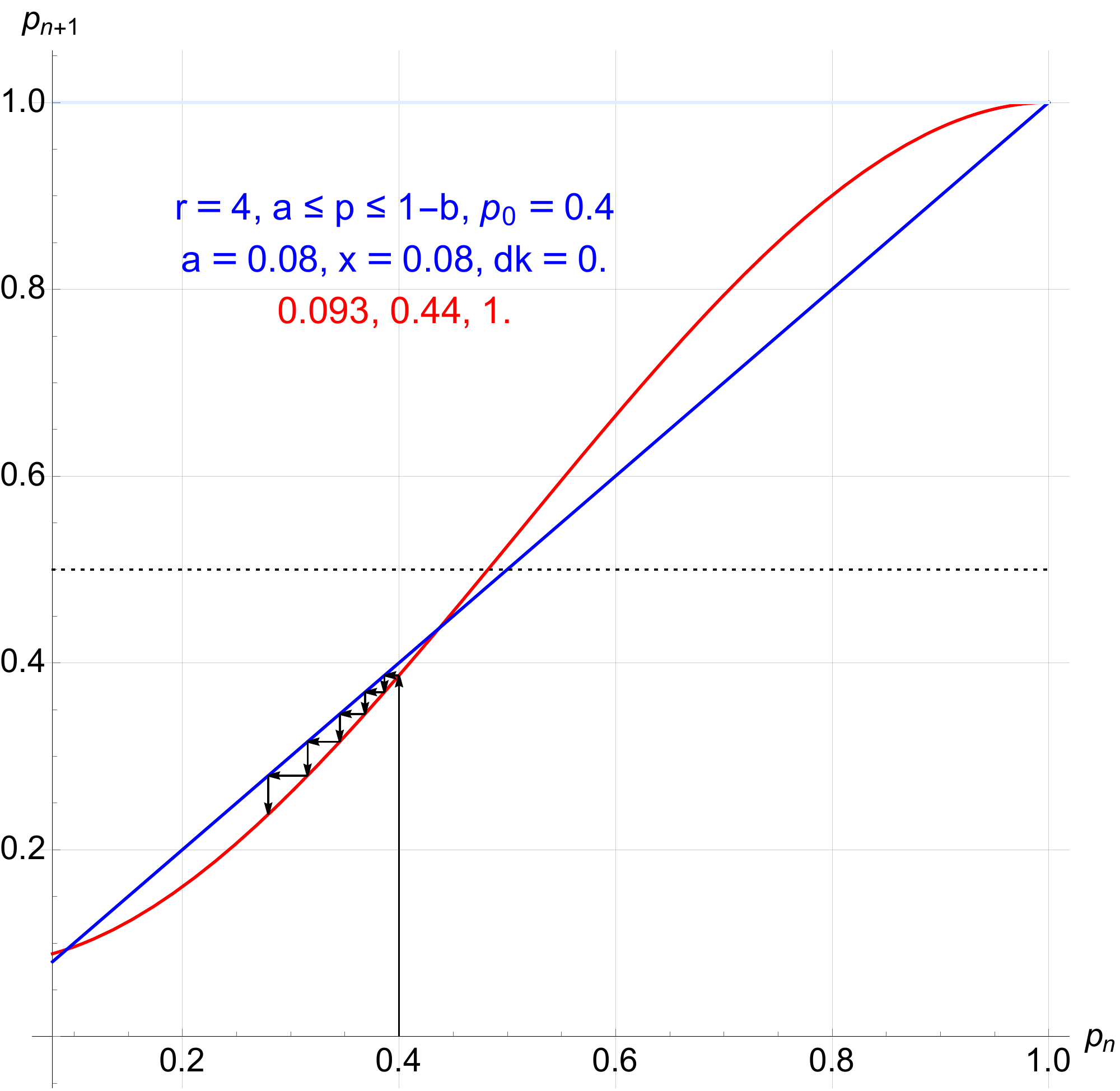}
\includegraphics[width=8cm,height=8cm]{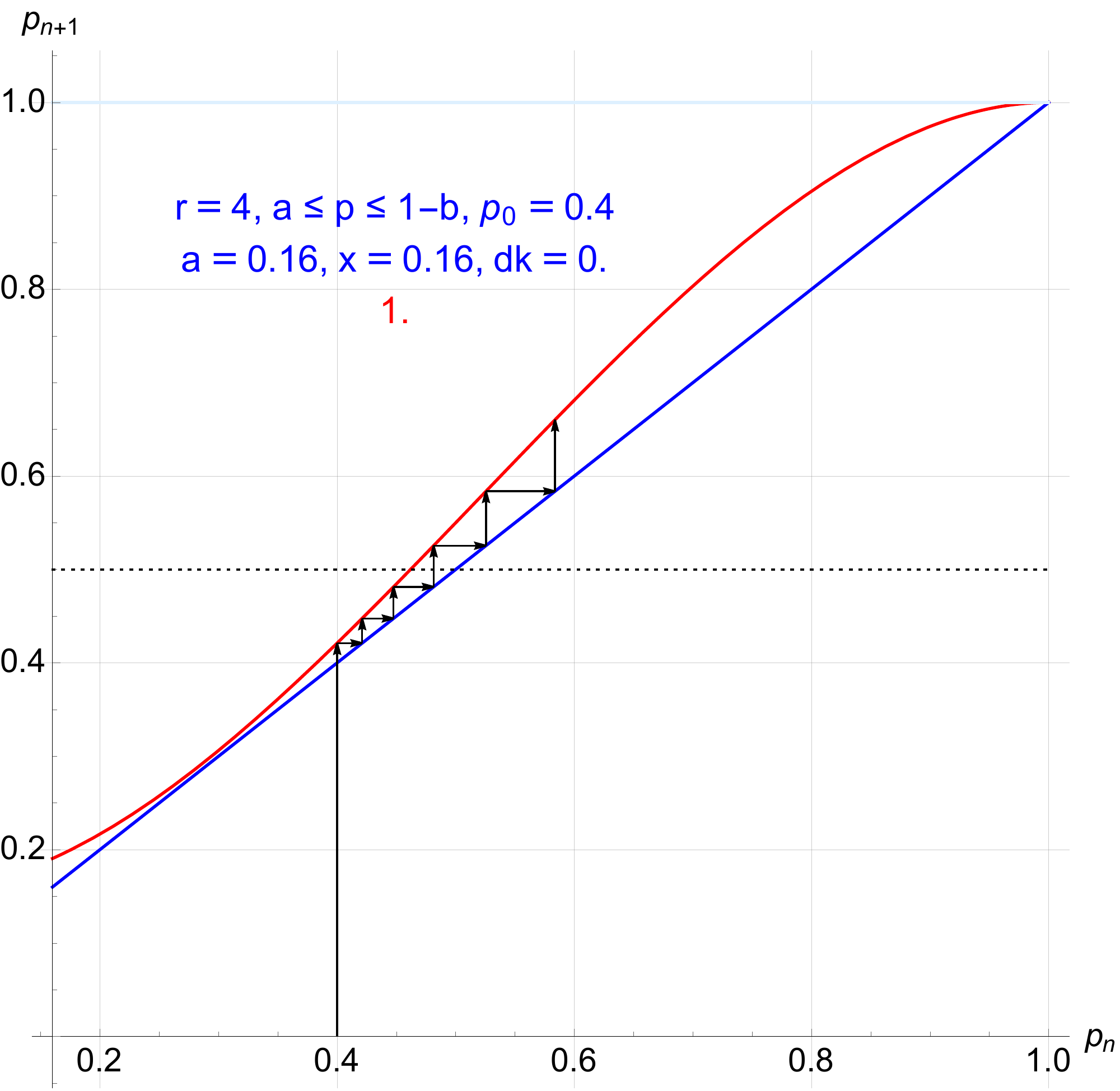}
\caption{Two different one sided inflexibles ($x=a \Leftrightarrow b=0$) with balanced prejudices $dk=0$ for group size 4. The arrows show the dynamics starting from $p_0=0.40$. Left: $a=0.08$ yields $p_A=1, p_t=0.44, p_B=0.093$. Right: $a=0.16$ yields $p_A=1$ making A always to win.
}
\label{o1}
\end{center}
\end{figure}

\begin{figure}
\begin{center}
\includegraphics[width=8cm,height=8cm]{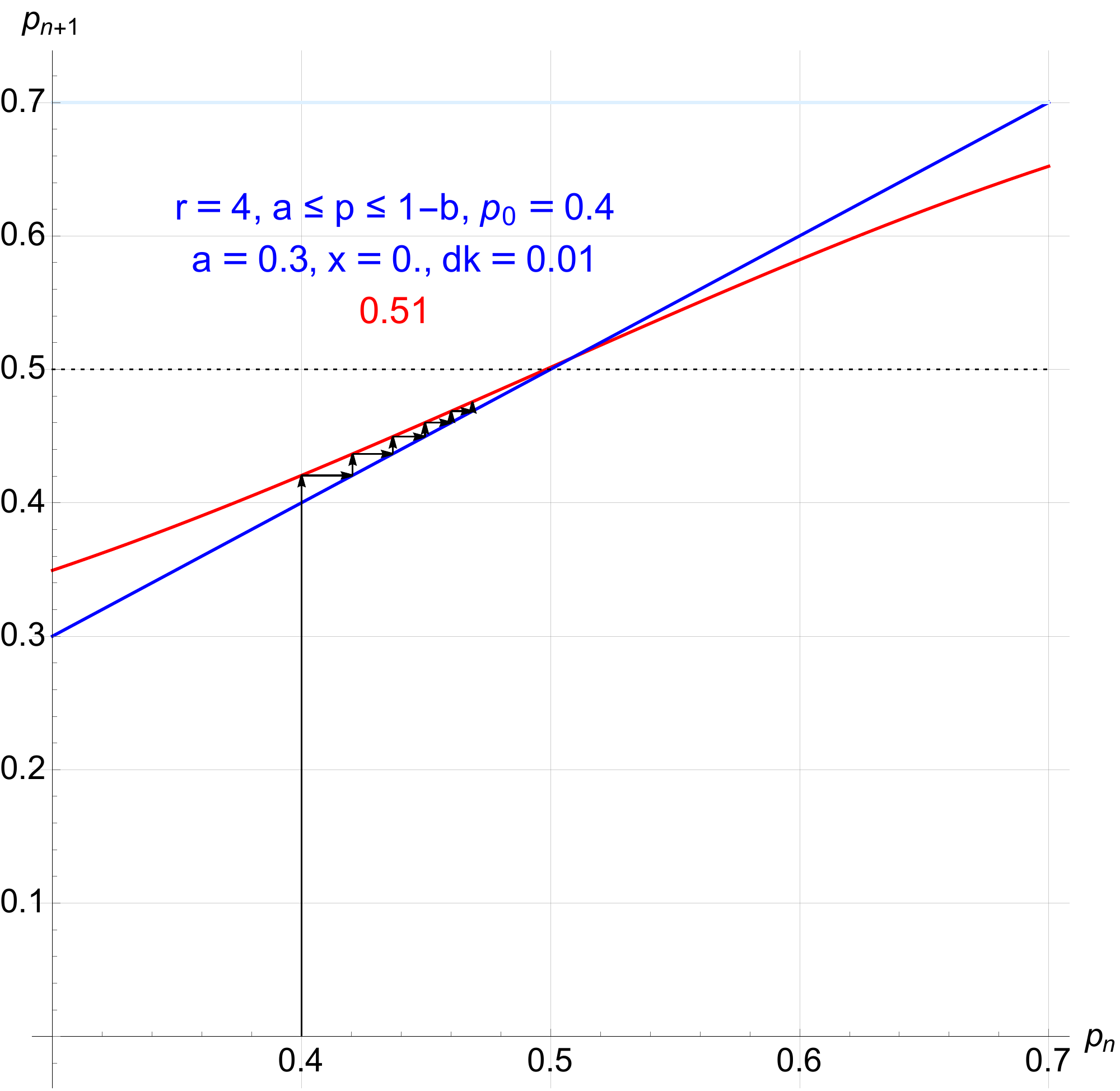}
\includegraphics[width=8cm,height=8cm]{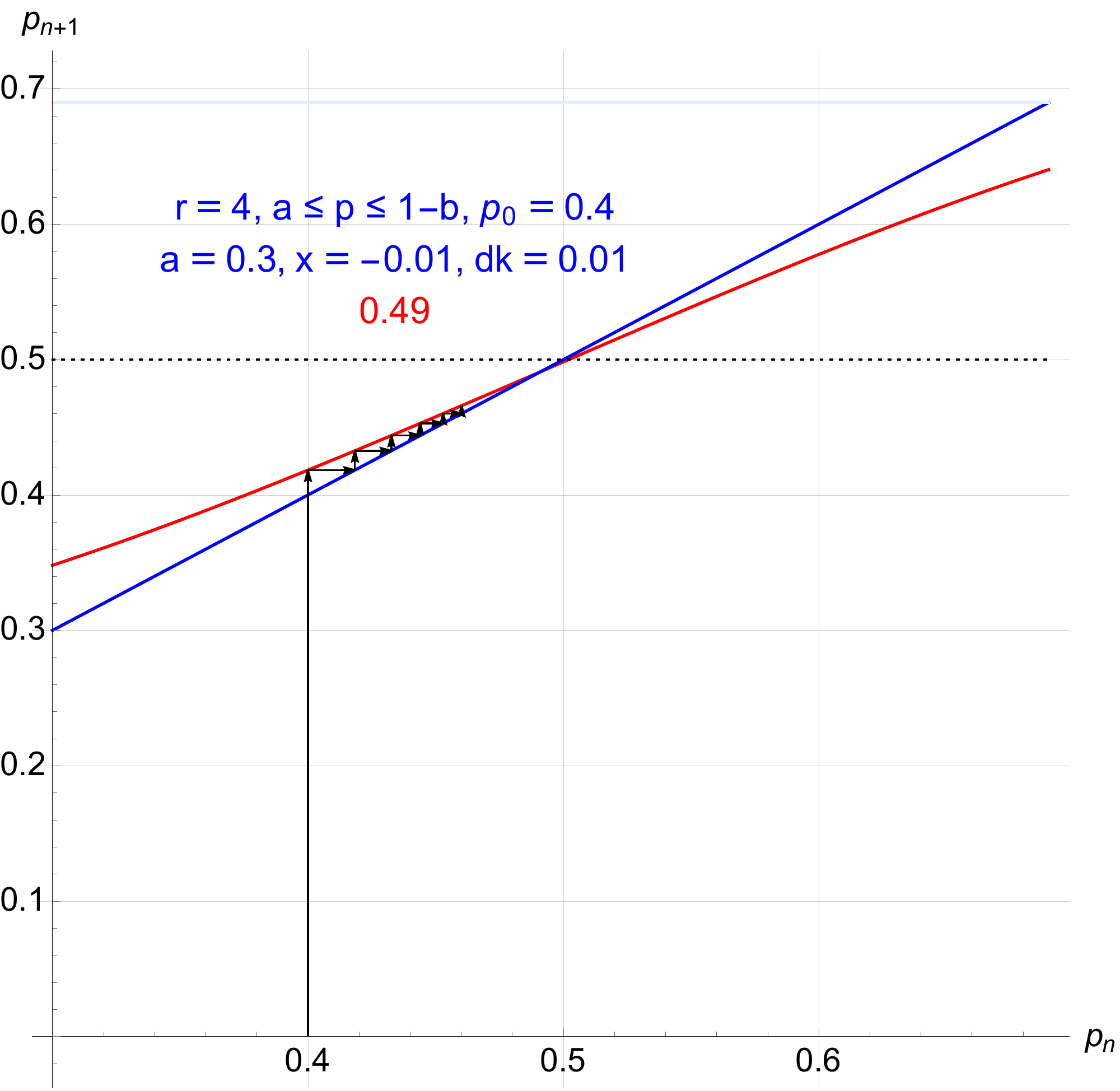}
\caption{Two different cases with $a=0.30$ and group size 4. The arrows show the dynamics starting from $p_0=0.40$. Right: $x=0, dk=0.01$ yields $p_s=0.51$ (A wins). Left: $x=-0.01, dk=0.01$ yields $p_s=0.49$  (A loses).
}
\label{o2}
\end{center}
\end{figure}

\begin{figure}
\begin{center}
\includegraphics[width=8cm,height=8cm]{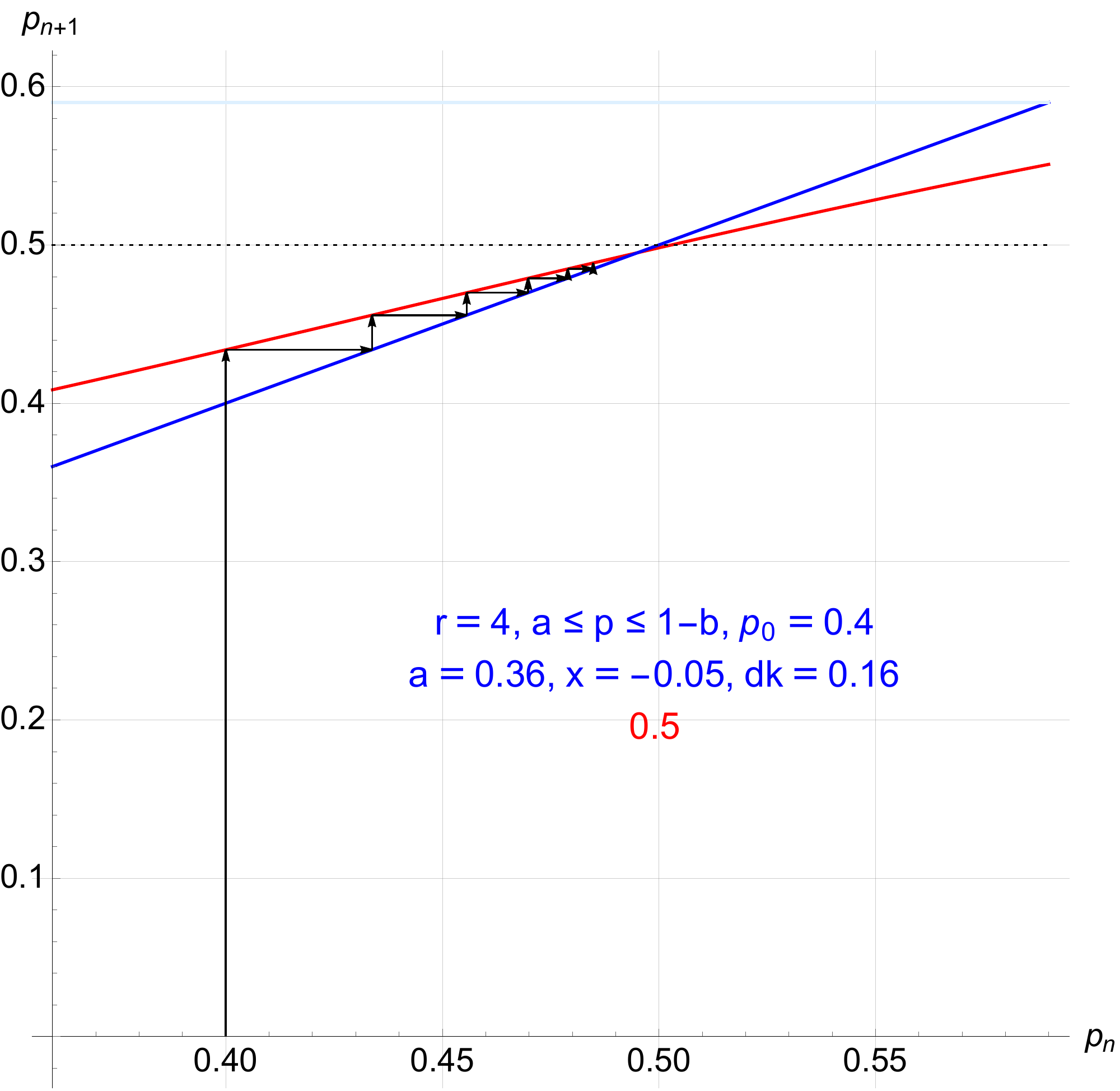}
\includegraphics[width=8cm,height=8cm]{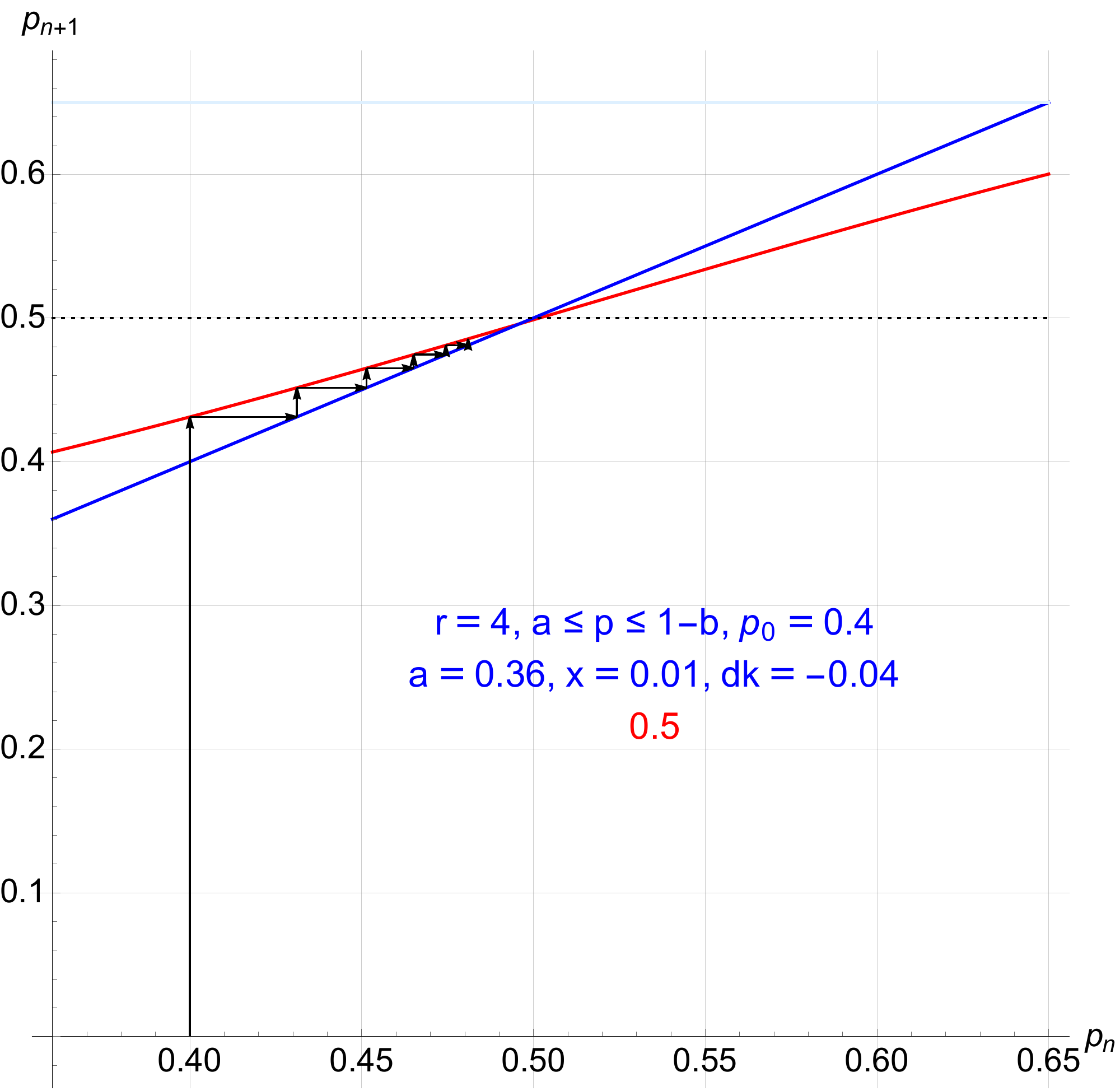}
\caption{Two different cases with $a=0.36$ and group size 4. The arrows show the dynamics starting from $p_0=0.40$. Left: $x=-0.05, dk=0.16$ yields $p_s=0.5$. Right: $x=0.01, dk=-0.04$ yields $p_s=0.50$. 
}
\label{o3}
\end{center}
\end{figure}

\begin{figure}
\begin{center}
\includegraphics[width=8cm,height=8cm]{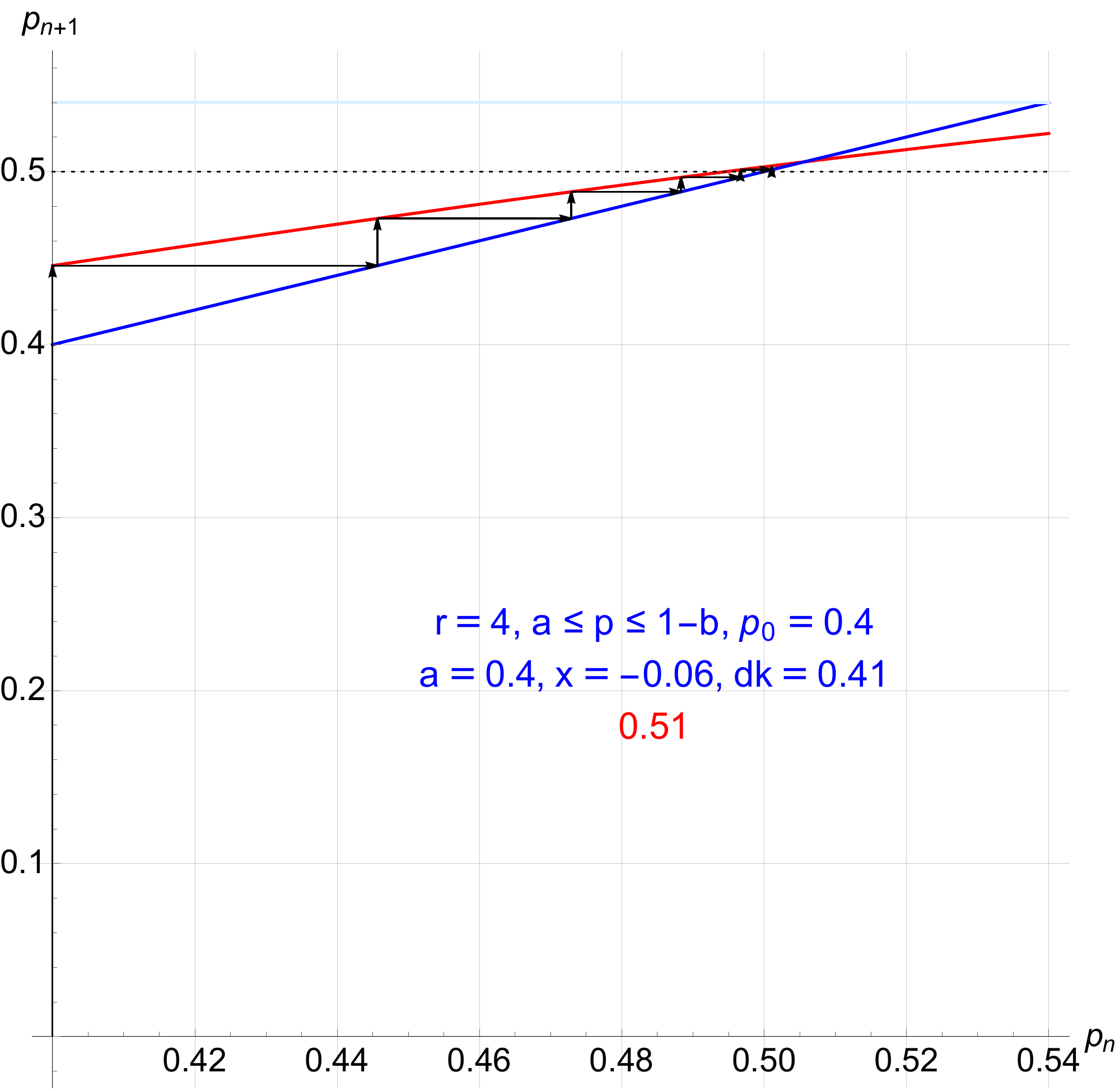}
\includegraphics[width=8cm,height=8cm]{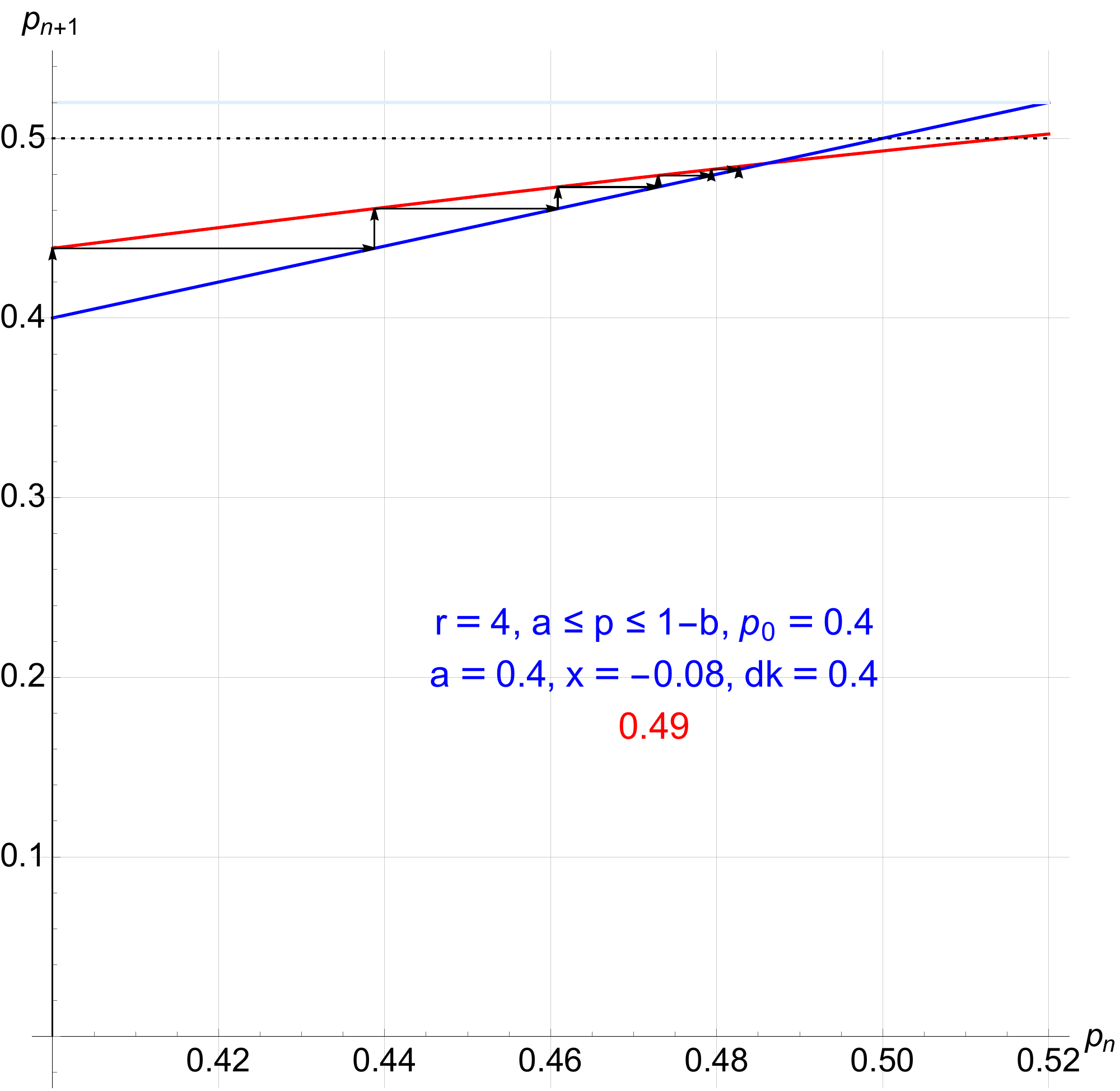}
\caption{Two different cases with $a=0.40$ and group size 4. The arrows show the dynamics starting from $p_0=0.40$. Left: $x=-0.06, dk=0.41$ yields $p_s=0.51$  (A wins). Right: $x=-0.08, dk=0.40$ yields $p_s=0.49$  (A loses).
}
\label{o4}
\end{center}
\end{figure}

To illustrate the large spectrum of unexpected dynamics produced by a combination of tie prejudice breaking and inflexible effects I show four  series of emblematic cases  with $r=4$ as a function of both $dk$ and $x$ in Figures (\ref{o1}, \ref{o2}, \ref{o3}, \ref{o4}). The cases with $p_s=\frac{1}{2}$ are chosen to show how tiny changes in either $dk$, $x$ or both will make the winner. This sensitivity is exhibited in the Figures with $p_s=0.51$ and $p_s=0.49$,

\begin{enumerate}
\item  
Figure (\ref{o1}) shows two one sided inflexibles  cases ($x=a \Leftrightarrow b=0$) with balanced prejudices $dk=0$ for group size 4. The arrows show the dynamics starting from $p_0=0.40$.  Left: $a=0.08$ yields $p_A=1, p_t=0.44, p_B=0.093$. Right: $a=0.16$ yields $p_A=1$ making A always to win.

\item 
Figure (\ref{o2}) shows two cases with $a=0.30$ and group size 4. The arrows show the dynamics starting from $p_0=0.40$. Right: $x=0, dk=0.01$ yields $p_s=0.51$ (A wins). Left: $x=-0.01, dk=0.01$ yields $p_s=0.49$  (A loses).

\item 
Figure (\ref{o3}) shows two different cases with $a=0.36$ and group size 4. The arrows show the dynamics starting from $p_0=0.40$. Right: $x=-0.05, dk=0.16$ yields $p_s=0.5$. Left: $x=0.01, dk=-0.04$ yields $p_s=0.50$. 

\item
Figure (\ref{o4}) shows two different cases with $a=0.40$ and group size 4. The arrows show the dynamics starting from $p_0=0.40$. Left: $x=-0.06, dk=0.41$ yields $p_s=0.51$  (A wins). Bottom left: $x=-0.08, dk=0.40$ yields $p_s=0.49$  (A loses).

\end{enumerate}

\section{The 2020 prediction:  winning strategies}

What to conclude from above results, which allow envisioning novel disturbing strategies to win a major political vote, including the 2020, November 3, American presidential election? Contrary to what could be a priori expected, to win a voting majority is not to convince a maximum of floaters. 

For each candidate, main instrumental keys appear to be twofold, focusing on both increasing the share of the naturally activated prejudices which are in tune with the candidate and producing the maximum number stubborn supporters. Then, the goal is to initiate a large number of small group informal discussions. Different approaches have been elaborated to optimize winning strategies \cite{mar1,mar2}.

On this basis, as seen from the series of Figures (\ref{o1}, \ref{o2}, \ref{o3}, \ref{o4}), what matters are both excesses of prejudices and inflexibles measured by respectively $dk$ and $x$. Even tiny excesses are instrumental to win. When both $dk>0$ and $x>0$ for the same candidate, according to the model, this candidate is almost sure to win the election in the corresponding state. When  $dk>0$ and $x<0$ or $dk<0$ and $x>0$ both effect compete making the outcome more uncertain.

\textbf{Stubbornness:} millions of American are either stubborn anti-Trump or stubborn pro-Trump and it is difficult to get solid figures of their respective proportions. However, the beginning of the campaign for the democrat nomination has disclosed very large gaps between Sanders and Biden supporters. 

This observation suggests that not every stubborn anti-Trump voter is a stubborn pro-Biden. The lack of total overlap between stubborn anti-Trump and stubborn pro-Biden voters produces some advantage for Trump with respect to his stubborn supporters, which can become significant in particular in a few swing states. 

Moreover, contrary to Biden, Trump has been very active on the ground with many meetings to rise the degree of stubbornness of his supporters.

\textbf{Prejudices:} among prejudices which could be activated during the 2020 campaign, I identified fear and personal stand, two major ones which can benefit to each candidate depending on respective psychological traits of agents. 

\begin{enumerate}
\item 
Fear about the other candidate election.

Both Biden and Trump have been waiving the fear about consequences of having the other candidate elected. Depending on the social groups, there exists a fear of Trump or a fear of Biden. 

However, the fear of Trump has been eroded in part since people know what he does. Thus, Trump generates more stubborn anti-Trump than fear. Moreover, his claim of “law and order” mitigates the fear of a second mandate by producing a sense of credible reassurance among some voters, who are afraid of the possible chaos and massive immigration, which could follow Biden election. 

In addition, the fear of Biden related to his supposed socialist project may be fueled from irrational phantasms among parts of the people as socialism is both unknown and rooted in the former cold war fears, in US history. 
\item 
Personal stance facing a danger.

While Trump handling of the Covid 19 epidemics has been penalizing him among many voters, his being infected and cured may compensate part of his loose in support. This point relies on the perception of what is the best personal stance for a president facing a danger. The preference for showing strength and being reckless (Trump) versus being cautious and careful (Biden) varies depending on the individual voter psychology, which can be decoupled from the voter political camp.
\end{enumerate}

\section{The 2020 prediction: the November winner}

From above results the winner in the 2020 November election will be the candidate who will succeed in getting more $dk>0$ and $x>0$ in a series of swing states to reach the majority of Electors. However, in each state the various proportions of respective inflexibles (stubborn pro-Biden, stubborn pro-Trump), and leading prejudices (fear ofTrump or Biden, reckless or cautious), are unknown. On this basis, I can only make rough estimates to determine winning or losing trends. 

From my perception and analysis, Trump will benefit from the fact that blind anti-Trump voters will not be automatically blind Biden voters although they much overlap. As seen above, tiny differences in $x$ can make the outcome. In addition, Trump has been more involved than Biden on the ground with meetings to motivate his supporters, creating more stubbornness. Trump recover from being infected by Covid 19 should also be at his benefit towards $dk>0$ in comparison with Biden precautionary attitude.

Accordingly, using the Galam model of opinion dynamics, my prediction is that Trump, being more likely to have advantage in both $dk>0$ and $x>0$ within sufficient swing states, will eventually win the 2020, November 3 presidential election \cite{zoom}.

\subsection{Hidden voting and hidden abstention}

Independently of the opinion dynamics per se, I evoke two additional mechanisms , which should be working at Trump benefit on the voting days, hidden voting and hidden abstention.
\begin{itemize}
\item Hidden voting

Some voters supporting Trump, do not dare standing openly for their choice, which can be perceived as controversial by their acquaintances. They will then vote for Trump without disclosing neither their voting intention nor their actual vote. While such hidden voters are not present in pro-Trump neighborhood, they could be present in anti-Trump neighborhoods. The phenomena is more likely to occur in favor of Trump than in favor of Biden.

\item Hidden abstention

Some former pro-Sanders voters may found too wide the gap between respective programs of Biden and Sanders. Therefore, although they are anti-Trump, they may tempted to abstain on the voting day. 

I had introduced the phenomenon under the name "unavowed abstention" for 2017 French Presidential election where M Le Pen was competing with E. Macron \cite{nyt}. Finally, the phenomenon did occur but surprisingly at Macron benefit \cite{una}. However, if hidden abstention takes place it will be only at Biden expense.

\end{itemize}

\section{Conclusion}

Using the Galam model of opinion dynamics, I was able to predict successfully the 2016 Trump victory, who had found a winning martingale, which could not be applied by Clinton. However, for the coming 2020 election, according to the same model, the instrumental winning quantities $dk$ and $x$  to ensure the victory, are this time available to both candidate. Biden and Trump can increase the stubbornness of their supporters and build up fear for the other candidate as well as promoting either being reckless or cautious. However, my rough estimates are that Trump has advantages along those features in sufficient swing states to make him likely to win in the November 3 election.

To conclude, I want to emphasize that my prediction is not about taking risk or becoming a kind of guru if it proves itself right. It is about a hard science approach to model political events. 
\begin{itemize}
\item If the prediction is successful, it will not mean the model is proved, but that the model does capture some mechanisms at work in the driving of opinion dynamics and more investigation is worth along its path.

\item It the prediction turns wrong, it will not mean the model has to be thrown away, but that I missed some important feature, which must be identified and added to it. It can be also due to a wrong estimates of  $dk$ and $x$ in some states.
\end{itemize}

I am not dealing here with a work in progress. It is a field in progress with numerous physicists all over the world working to establish sociophysics on solid ground \cite{emo, janu1,mal,sab,bar,vil,janu2,kauf,kat,poin,bos}. The associated issues at stake are huge for our future.

I must also stress that the strategies elaborated to win the election are ethically questionable and not glorious. However, if these strategies turn to be valid, it will be an unfortunate but meaningful discovery of the modeling, which may bring into question our view and use of public opinion. These results enlighten the need to unveil the laws governing our collective behavior to avoid being trapped by our archaism in today world of connected people.

\section*{ Note added in Revision}

My prediction failed and as anticipated in the conclusion, it deserves an analysis to single out what went wrong in the making of the prediction. Two possibilities exist about a failure, either it is a significant failure with a massive blue wave as predicted by many polls or a light failure with Biden winning at the edge. The first case would require revisiting the basic elements of the model with some relevant ingredient being missed. On the contrary, the second case would be coherent with the basic elements of the model with error in the estimate of respective proportions of inflexibles and prejudice effects. Therefore, to  address the issue, I need to answer the following questions:

\begin{enumerate}
\item  Did Biden won with a substantial advantage or did he win at the edge?

First, focusing on the national vote with an excess of 7 millions votes for Biden with about 81 millions ballots against about 74 millions for Trump is misleading since the American presidential electoral system is quite specific and rather peculiar with the president being elected by the Elector college \cite{US}.

Second, looking at the Electoral vote would lean to assess a major victory for Biden who got 74 more votes with 306 votes against 232 for Trump. However, the assessment must be tempered. Looking at the Electoral vote in more details in three swing states, Arizona, Georgia and Wisconsin reveals that Biden won  at the edge.

The excess of ballots in favor of Biden against Trump, amounts respectively in these three states to 10 457,  11 779 and 20 682, which represent 0.31\%, 0.24\% and 0.63\% of the total ballots. All three figures are very tiny, which makes Biden victory very tight. 

To illustrate this statement, I point out that for instance the hypothetical scenario having 5889+1, 10341+1, 5228+1 Biden voters shifting their votes for Trump in Arizona, Georgia and Wisconsin would gives the 11, 16 and 10 associated Electors to Trump, creating a tie with 269 Electors to each candidate. It would represent 0.18\% ballots over the total of 11 685 327 ballots (3 387 326, 4 999 960, 3 298 041) in the three states.
 
And here comes another feature of the American electoral system with the 12th Amendment of the constitution. At a tie in the Electoral college, the House of Representatives elects the president with every state getting one vote \cite{tie1}. Republicans had a majority of 52 till January 3, a majority of 51 till January 18, and a tie from January 20. The President validation being held on January 10, Trump would have then been elected \cite{tie2}.
 
I developed above scenario to emphasize that while Biden victory is clear, it is also tight.

\item Is the prediction error related to the model itself?

My above illustration, which would have lead to Trump victory, shows that my prediction failed very close to the actual outcome, contrary to most poll predictions, which were predicting a large victory for Biden. This fact enlightens my underlining that tiny differences will make big outcomes as in "The cases with $p_s=\frac{1}{2}$ are chosen to show how tiny changes in either $dk$, $x$ or both will make the winner. This sensitivity is exhibited in the Figures with $p_s=0.51$ and $p_s=0.49$,"

\item What is the source of the error?

I have assumed Trump would have a little more inflexibles than Biden while in reality it has been the opposite with Biden having a little more inflexibles than Trump. In terms of the model, that is a minor change although the associated outcome is highly significant, along to what is mentioned in the manuscript about the high impact of tiny differences. Moreover, the fact that Trump cured from Covid 19 did not impact the prejudice effect at his benefit. 

As one referee pointed out, my rough estimates for $dk$ and $x$ were obtained using personal beliefs and impressions, which a priori is not sound with respect to a scientific approach. As quoted from the referee,  "The author should have take into account real data to estimate the value of the parameters used in the model...  this is a very weak point of the paper".

Although the referee's statement makes sense, here, it is out of context because such data are not available. They do not exist at the moment. In addition, my estimates happened to be rather close to the reality as seen from above illustration. Indeed, I would say that using my personal beliefs and impressions to give an estimate of the model's parameters, is not a weak point of the paper, but instead the result of a weak point of the current knowledge about voter psychologies with the absence of specific data connected to the model parameters. For the future, the design and implementation of schemes and procedures to make those estimates is a major key to turn the model into a solid predictive tool. However, such a task, which is beyond my reach, will be a real challenge, given the repeated failures of polls for both 2016 and 2020 US elections.

\item Why the absence of reliable data did not preclude my successful 2016 prediction?
 
As underlined in the paper, for the 2016 election the tie breaking prejudice was activated only at Trump benefit making the outcome more robust with respect to a solid estimate of the parameters. However, in 2020 election, Biden and Trump were equally competing on both grounds of inflexibles and tie breaking prejudice. This is why I considered the difference in respective proportions of inflexibles and  the departure from fifty percent for prejudice. Such a situation makes estimates of the parameters more tricky and more sensitive to the outcome as shown with the failure of the prediction. The competition in the swing states was predicted to be around fifty-fifty as shown in several of my Figures.

\end{enumerate}

In the paper I also mentioned hidden voting and hidden abstention as two possible features which would benefit to Trump independently of the dynamics of opinion. Hidden voting seems to have been significant when comparing the polls prediction and the very high turnover for Trump with more than 74 millions voters. In contrast, hidden abstention did not occur as seen from the extremely high turnover for Biden with more that 81 millions voters.
 
\section*{Acknowledgement}

I would like to thank late Dietrich Stauffer for our numerous lively and exciting discussions during several decades. He has been an outstanding physicist and a friend who has been always available to respond to my many comments with sound arguments and an amazing sense of humor. I am convinced he would have loved a paper with the author arguing that its wrong prediction was "almost right". I miss him.



\begin{thebibliography}{99}


\bibitem{bra} R. Brazil, 
The physics of public opinion, 
Physics World, January issue (2020).

\bibitem{noor} H. Noorazar, 
Recent advances in opinion propagation dynamics: A 2020 Survey, 
The Eur. Phys. J. Plus {\bf 135}, Article number: 521 (2020).

\bibitem{fran} F. Schweitzer, 
Sociophysics, Physics Today {\bf 71}, 
40--47 (2018).

\bibitem{sprin} S. Galam, 
{\em Sociophysics: A physicist's modeling of psycho-political phenomena}, 
Springer, New York (2012).

\bibitem{cast} C. Castellano, S. Fortunato and V. Loreto, 
Statistical physics of social dynamics, 
Rev. Mod. Phys. {\bf 81}, 591--646 (2009).

\bibitem{tru} S. Galam, 
The Trump phenomenon, an explanation from sociophysics, arXiv:1609.03933 (2016);
Int. J. Mod Phys. B {\bf 31}, 1742015(17pp) (2017).

\bibitem{lich} https://www.nytimes.com/2020/08/05/opinion/2020-election-prediction-allan-lichtman.html

\bibitem{pol} https://ig.ft.com/us-election-2020/

\bibitem{frans} S. Galam and F. Jacobs, 
The role of inflexible minorities in the breaking of democratic opinion dynamics,
Physica A {\bf 381}, 366--376  (2007).

\bibitem{mas} A. Cardillo and N. Masuda,
Critical mass effect in evolutionary games triggered by zealots,
Phys. Rev. Res.2, 023305 (2020).

\bibitem{JP18} J. S. Juul  and M. A. Porter, 
Hipsters on networks: How a minority group of Individuals can lead to an anti-establishment majority, 
Phys. Rev. E 99, 022313 (2019).

\bibitem{PSL12} W. Pickering, B. K. Szymanski and C. Lim, 
Analysis of the high dimensional naming game with committed minorities, 
Phys. Rev. E {\bf 93}, 052311(9pp)  (2016). 

\bibitem{BRG16} K. Burghardt, W. Rand and M. Girvan, 
Competing opinions and stubbornness: connecting models to data,  
 Phys. Rev. E {\bf 93}, 032305(14pp) (2016).

\bibitem{CO15} N. Crokidakis and P. M. C. de Oliveira, 
Inflexibility and independence: Phase transitions in the majority-rule model, 
Phys. Rev. E {\bf 92}, 062122(9pp) (2015).

\bibitem{Mo15} M. Mobilia, 
Nonlinear q-voter model with inflexible zealots, 
Phys. Rev. E {\bf 92}, 012803(11pp) (2015).

\bibitem{MG13} A. Martins and S. Galam, 
Building up of individual inflexibility in opinion dynamics,
Phys. Rev. E {\bf 87}, 042807(8pp) (2013). 

\bibitem{pub} S. Galam, 
Public debates driven by incomplete scientific data: The cases of evolution theory, global warming and H1N1 pandemic influenza,
Physica A 389, 3619-363 (2010).

\bibitem{pair} S. Galam, 
Collective beliefs versus individual inflexibility: The unavoidable biases of a public debate, 
Physica A 390, 3036-3054  (2011).

\bibitem{hete} S. Galam, 
Heterogeneous beliefs, segregation, and extremism in the making of public opinions,  
Phys. Rev. E {\bf 71},  046123(5pp) (2005).

\bibitem{uni} S. Galam and T. Cheon, 
Tipping points in opinion dynamics: a universal formula in five dimensions, 
Frontiers in physics, 8, 446 (2020).

\bibitem{mar1}  M.A. Javarone, Network Strategies in the Election Campaigns , JSTAT, V2014 – P08013 (2014).

\bibitem{mar2} M.A. Javarone and T. Squartini, Conformism-driven phases of opinion formation on heterogeneous networks: The q-voter model case. JSTAT, V2015, P10002 (2015).

\bibitem{zoom} S. Galam, 
Will Trump win again in the Nov 4 (French time) election? 
https://www.youtube.com/watch?v=r\_PDot-b8eE

\bibitem{nyt} S. Galam, 
Abstention, France’s last temptation, 
New York Times, May 5 (2017).

\bibitem{una} S. Galam, 
Unavowed abstention can overturn poll predictions, 
Frontiers in Physics, 00024 (2018).

\bibitem{emo} D. Tsarev, A. Trofimova, A. Alodjants and A. Khrennikov,  Phase transitions, collective emotions and decision-making problem in heterogeneous social systems. Sci Rep 9, 18039 (2019).

\bibitem{janu1} J. A.Ho{\l}yst, K. Kacperski and F. Schweitzer, Phase transitions in social impact models of opinion formation,
Physics A 285, 199-210 (2000).

\bibitem{mal} A. Kowalska-Stycze\'n, K. Malarz,
Noise induced unanimity and disorder in opinion,
formation. PLoS ONE 15(7): e0235313 (2020).

\bibitem{sab} D. Sabin-Miller and D. M. Abrams,
When pull turns to shove: A continuous-time model for opinion dynamics,
Phys. Rev. Res. 2, 043001 (2020).

\bibitem{bar} A. F. Siegenfeld and Y. Bar-Yam,
Negative representation and instability in democratic elections,
Nature Physics volume 16, 186–190 (2020).

\bibitem{vil} A. L. M. Vilela, L. F. C. Pereira, L. Dias, H E. Stanley and L. R. da Silva,
Majority-vote model with limited visibility: an investigation into filter bubbles,
Physica A  563, 125450 (2021).

\bibitem{janu2} {\L}. G. Gajewski, J. Sienkiewicz and J. A. Ho{\l}yst, Bifurcations and catastrophes in temporal bi-layer model of echo chambers and polarisation, arXiv:2101.03430v1 (2021).


\bibitem{kauf} M. Kaufman, H. T. Diep and S. Kaufman,
Sociophysics Analysis of Multi-Group Conflicts 
Entropy 22(2), 214 (2020)

\bibitem{kat} B. Nowak and K.  Sznajd-Weron,
Homogeneous Symmetrical Threshold Model with Nonconformity: Independence versus Anticonformity,
 Complexity vol. 2019, ID 5150825, 1-14 (2019).
 
 
 \bibitem{poin} A. Poindron, A general model of binary opinions updating, Mathematical Social Sciences 109,  52-76 (2020).

\bibitem{bos} G. Boschi, C. Cammarota and R. Kühn, Opinion dynamics with emergent collective memory: A society shaped by its own past, Physica A 558, 124909 (2020).


\bibitem{US} $https://en.wikipedia.org/wiki/2020\_United\_States\_elections$

\bibitem{tie1} $https://www.brookings.edu/blog/fixgov/2020/12/09/the-electoral-college-is-a-ticking-time-bomb/$

\bibitem{tie2} $https://en.wikipedia.org/wiki/117th\_United\_States\_Congress$

\end{thebibliography}
\end{document}